# Field-free spin-orbit-torque switching of perpendicular magnetization aided by uniaxial shape anisotropy

Zhaohao Wang[1,2], Zuwei Li[1,2], Min Wang[1], Bi Wu[1], Daoqian Zhu[1] and Weisheng Zhao[1,2]

[1] School of Microelectronics, Beijing Advanced Innovation Center for Big Data and Brain Computing, Fert Beijing Research Institute, Beihang University, Beijing, 100191, China
[2] Beihang-Goertek Joint Microelectronics Institute, Qingdao Research Institute, Beihang University, Qingdao 266100, China

E-mail: zhaohao.wang@buaa.edu.cn; weisheng.zhao@buaa.edu.cn

**Abstract**

It has been demonstrated that the switching of perpendicular magnetization can be achieved with spin orbit torque (SOT) at an ultrafast speed and low energy consumption. However, to make the switching deterministic, an undesirable magnetic field or unconventional device geometry is required to break the structure inverse symmetry. Here we propose a novel scheme for SOT-induced field-free deterministic switching of perpendicular magnetization. The proposed scheme can be implemented in a simple magnetic tunnel junction (MTJ) /heavy-metal system, without the need of complicated device structure. The perpendicular-anisotropy MTJ is patterned into elliptical shape and misaligned with the axis of the heavy metal, so that the uniaxial shape anisotropy aids the magnetization switching. Furthermore, unlike the conventional switching scheme where the switched final magnetization state is dependent on the direction of the applied current, in our scheme the bipolar switching is implemented by choosing different current paths, which offers a new freedom for developing novel spintronics memories or logic devices. Through the macrospin simulation, we show that a wide operation window of the applied current pulse can be obtained in the proposed scheme. The precise control of pulse amplitude or pulse duration is not required. The influences of key parameters such as damping constant and field-like torque strength are discussed as well.

## 1. Introduction

High-speed and energy-efficient switching mechanism of perpendicular magnetization is strongly desired in spintronics applications, as it could benefit the write performance of magnetic random access memories and spin logic devices [1-3]. Currently the most widely used mechanism is spin transfer torque (STT) [4-5], which, however, is suffering from several intrinsic drawbacks. For instance, the STT switching speed is limited by the incubation delay. In addition, the STT-based device is easily impaired as the switching current directly flows through a tunnel barrier. Recently an alternative mechanism called spin orbit torque (SOT) has been paid much attention as it can solve the above-mentioned drawbacks of the STT [6-12].

Typically, the SOT is induced in a non-magnetic/ferromagnetic/insulator (NM/FM/I) heterostructure, where the NM layer may be heavy-metal [6-8], antiferromagnet [13, 14] or topological insulator [15-17] with strong spin-orbit coupling (SOC). A charge current passing through the NM layer generates the accumulation of spins and thereby exerts a torque (i.e. SOT) to switch the magnetization of FM layer. Compared with the STT, the current required by the SOT flows through the NM layer instead of the tunnel barrier, thus the write endurance of the device is enhanced. Furthermore, the SOT can induce ultrafast switching (e.g. sub-nanosecond) of perpendicular magnetization by eliminating time-consuming precessional motion. Such a speed cannot be easily implemented with the conventional STT since the STT is triggered by the thermal fluctuation.

However, the sole SOT drives the magnetization to in-plane direction, thus the magnetization will relax upwards or downwards at the equal probability after the SOT is turned off. In other words, the switching of perpendicular magnetization induced by sole SOT is non-deterministic. Generally, an external magnetic field is applied to break the structure inverse symmetry so that the SOT-induced switching could be deterministic. Unfortunately the use of magnetic field is undesirable in the practical products, thus alternative solutions need to be discovered. In this context, numerous mechanisms of field-free SOT switching of perpendicular magnetization has been proposed, including breaking the lateral inversion symmetry [18], engineering a tilted anisotropy [19], inducing the exchange bias with antiferromagnet [13, 14, 20-21], and so on [22-33]. Nevertheless, these solutions require either uncommon fabrication process or precise control of pulse duration.

In this work, we propose a novel mechanism for the deterministic switching of perpendicular magnetization with field-free SOT. The key idea is combining the uniaxial shape anisotropy with the orthgonal SOT, by patterning a MTJ into elliptical shape and misaligning it with the axis of the heavy metal. The proposed mechanism does not require the complicated device structure. The perpendicular magnetization can be switched to the opposite states depending on the current paths, regardless of the current directions. The amplitude of the applied current pulse needs to be larger than a threshold, but the accurate duration is not necessary. The robustness of the proposed switching mechanism is validated by investigating the switching probability under the thermal fluctuation. The influences of damping constant and field-like torque are discussed as well.

## 2. Simulation models

The typical structure of the studied device is illustrated in Figure 1(a), where a magnetic tunnel junction (MTJ) is deposited above a heavy metal (HM) layer. The MTJ is composed of a tunnel barrier (TB) sandwiched between two FM layers, which are called free layer (FL) and pinned layer (PL), respectively. The FL (PL) has switchable (pinned) perpendicular magnetization. The tunneling resistance of the MTJ is high (low) if the magnetization of the FL is anti-parallel (parallel) to that of the PL. The cross section of the MTJ is patterned into elliptical shape so that the energy landscape can be modified by the demagnetization effect. The long-axis of the elliptical cross section is intentionally tilted by an angle $\phi$ from y-axis of the heavy metal.

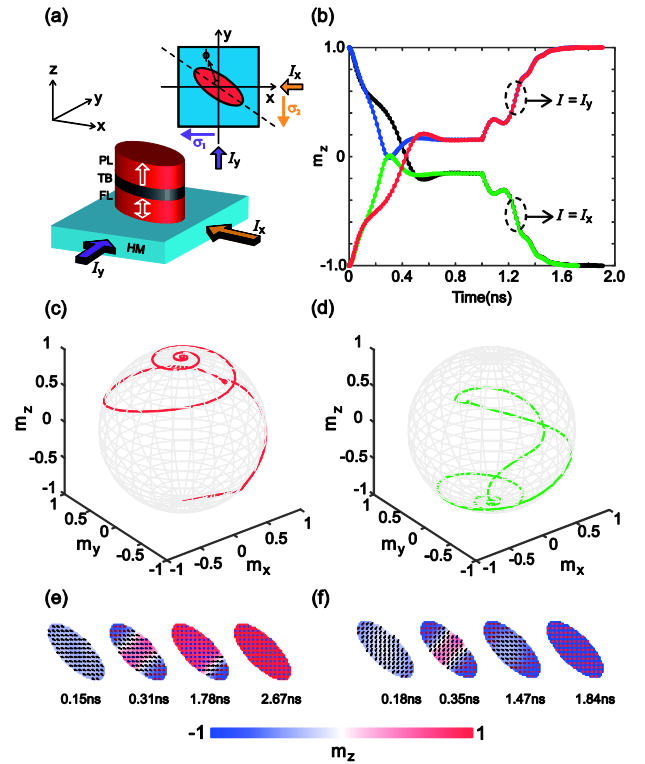

**Figure 1.** (a) Schematic of a typical device studied in this work. Inset: top view of the device. (b) Time evolution of z-component magnetization ($m_z$) under various initial states, current directions, and current paths. The current is applied along x-axis for the green and black curves, or along y-axis for the red and blue curves. (c)-(d) Trajectories of the magnetization when (c) $m_z$ is switched from -1 to 1 by applying $I_y$, and (d) $m_z$ remains at -1 by applying $I_x$. (e)-(f) Time-dependent micromagnetic configurations for (e) $m_z$ is switched from -1 to 1 by applying $I_y$, and (f) $m_z$ remains at -1 by applying $I_x$. In (b)-(f), the duration of the applied current pulse is set to 1 ns.

To deterministically switch the magnetization of the FL in Figure 1(a), we propose a novel scheme where a charge current could be applied to HM layer along y-axis or x-axis. In other words, there are two options for the paths of write current. The charge current ($I_y$) along y-axis generates the accumulation of x-axis-polarized spins ($\sigma_1 // I_y \times z$), which induces the SOT driving the magnetization towards x-axis direction. Meanwhile the tilted elliptical MTJ plane leads to a y-axis component of the shape anisotropy field, which is combined with the SOT to contribute the deterministic switching of the perpendicular magnetization. To perform the opposite switching, the path of the charge current passing the heavy metal layer is changed to x-axis ($I_x$). In this case, the relationship between the accumulated spins ($\sigma_2 // I_x \times z$) and the shape anisotropy field is inversed. As a result, the magnetization of the FL is switched to the opposite state. Therefore, in the proposed scheme the final state of the switched magnetization is dependent on the current path, which is entirely different from the existing schemes.

TABLE I. Parameters for the simulation.

| Symbol | Parameter | Default value |
|---|---|---|
| - | MTJ area | 40 nm × 100 nm × $\pi/4$ |
| - | Magnetic anisotropy | $9.5 \times 10^5 \, J/m^3$ |
| $M_s$ | Saturation Magnetization | $1.2 \times 10^6 \, A/m$ |
| $t_F$ | Free layer thickness | 1 nm |
| $\Theta_{SHE}$ | Spin Hall angle | 0.3 |

The magnetization dynamics of the FL is mathematically described by the following Landau-Lifshitz-Gilbert (LLG) equation

$$\frac{\partial \mathbf{m}}{\partial t} = -\gamma\mu_0 \mathbf{m} \times \mathbf{H}_{eff} + \alpha \mathbf{m} \times \frac{\partial \mathbf{m}}{\partial t} - J\xi\lambda_{DL}\mathbf{m} \times (\mathbf{m} \times \boldsymbol{\sigma}) \\ - J\xi\lambda_{FL}\mathbf{m} \times \boldsymbol{\sigma} \quad (1)$$

where $\gamma$ is the gyromagnetic ratio, $\mu_0$ is the vacuum magnetic permeability, $\alpha$ is the damping constant, $J$ is the applied current density, **m** is the unit vector of the magnetization, $H_{eff}$ is the effective field including the contribution of uniaxial anisotropy field and demagnetization field. The third and fourth terms at the right side are damping-like SOT and field-like SOT, respectively. $\lambda_{DL}$ and $\lambda_{FL}$ are factors reflecting the magnitudes of damping-like SOT and field-like SOT, respectively. $\xi = \gamma\hbar\Theta_{SHE}/(2et_FM_s)$ is a device-dependent parameter with $\hbar$ the reduced Planck constant and $e$ the electron charge. The other parameters and their default values are listed in Table I, which is consistent with the state-of-the-art technology.

In the preliminary study, the damping constant is set to an adequately large value ($\alpha = 0.2$). The tilted angle $\phi$ is set to $45°$ for obtaining the symmetry between $+\mathbf{z} \rightarrow -\mathbf{z}$ and $-\mathbf{z} \rightarrow +\mathbf{z}$ switching events. It is important to mention that the proposed mechanism can also work in the case of smaller damping constant or $\phi \neq 45°$, which will be explained later. For clearly analyzing the role of the SOT, we firstly neglect the thermal fluctuation and field-like SOT (i.e. $\lambda_{FL} = 0$), whose contribution will be discussed in the later paragraphs.

## 3. Results and discussion

Figure 1(b) shows typical macrospin simulation results of time-dependent z-component magnetization ($m_z$) under 8 possible combinations of {initial state ($\pm m_0$), current direction ($\pm I$), current path ($I_x$ or $I_y$)}. It is worth mentioning that only 4 curves are identified in Figure 1(b), because the change of current direction has no effect on the time evolution of $m_z$ (i.e. $+I$ and $-I$ correspond to the same curve in Figure 1(b)). The fundamental reason is: changing the current direction only induces the inversion of the sign of spin polarization vectors ($+\boldsymbol{\sigma}$ or $-\boldsymbol{\sigma}$), but $+\boldsymbol{\sigma}$ and $-\boldsymbol{\sigma}$ are symmetrical with respect to $m_z$. Thus the direction of the write current can be fixed if the proposed device is used for designing the MRAM, which helps in eliminating the source degeneration issue of the access transistor [34, 35]. Another two conclusions can be drawn from Figure 1(b). First, the final state is only dependent on the current path, in agreement with the above theoretical analysis. Second, the tilted angle $\phi = 45°$ induces mirror symmetry of spin torque between the cases $\{+m_0, I_x\}$ and $\{-m_0, I_y\}$, therefore the red and black curves (or, the green and blue curves) are symmetrical with respect to the line $m_z = 0$. Figure 1(c)-(d) show the typical trajectories of the magnetization dynamics. As can be seen, the switching is quickly activated without the incubation delay.

Overall 4 cases need to be considered while discussing the magnetization switching: i) $m_z$ is switched from $-1$ to $+1$ by applying $I_y$; ii) $m_z$ remains at $-1$ by applying $I_x$; iii) $m_z$ is switched from $+1$ to $-1$ by applying $I_x$; iv) $m_z$ remains at $+1$ by applying $I_y$. The magnetization dynamics of case i) or ii) is equivalent to that of case iii) or iv) according to the above analysis (see Figure 1(b)). Therefore in the following we only study the cases i) and ii), unless otherwise stated.

The above switching mechanism is also validated by micromagnetic simulation with OOMMF package, as shown in Figure 1(e)-(f). Compared with the macrospin simulation, where the magnetization is coherently switched, micromagnetic simulation shows multi-domain characteristic during the switching process. Despite this difference, the deterministic switching is unambiguously validated in both simulations. It is worth emphasizing that the required switching current density in the micromagnetic model is smaller than that in the macrospin model, since in the micromagnetic model the switching is activated by the local domain nucleation overcoming a lower energy barrier. Furthermore, the reported experimental results of the critical SOT current density is usually much smaller than the values predicted by macrospin or micromagnetic models due to various possible factors [8, 36-38]. Thus we argue that the switching current density shown in this work promises to be significantly reduced in the practical applications.

To evaluate the robustness of the proposed SOT mechanism, we study the switching and non-switching probabilities as a function of current density and pulse duration by considering the thermal fluctuation. The effect of thermal fluctuation is modelled by a Langevin stochastic field [39]. The phase diagrams of switching probability for cases i) and non-switching probability for case ii) are shown in Figure 2. As can be seen, when the current density and pulse duration are both large enough, the switching probability (Figure 2(a)) and non-switching probability (Figure 2(b)) would be almost 1. Note that in the left region of Figure 2(b) the non-switching probability is also 1, which is easy to understand since in this region the current is so small that the $m_z$ is slightly affected (i.e. remains at $-1$ and no switching occurs). The minimum pulse duration required by robust switching is as short as < 500 ps, which means that the effect of the SOT saturates at an ultrafast speed, in agreement with Figure 1(b). Once saturation, the magnetization fluctuates slightly around a stable position (denoted as $m_{stab}$) if the thermal noise is considered. After turning off the current, the magnetization would relax to $+z$ or $-z$ (final state) under the actions of $H_{eff}$ and damping torque. It is found that $m_{stab}$ is insensitive to the pulse duration due to the above-mentioned SOT saturation effect. This characteristic leads to a wide operation window of the applied current pulse. In comparison, the previously reported field-free SOT switching mechanisms require relatively strict control of current pulse [25, 31].

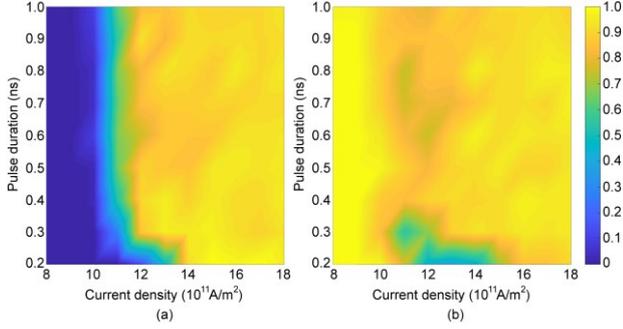

**Figure 2.** Phase diagrams of switching and non-switching probabilities as a function of current density and pulse duration. The data are obtained from 100 trials. The simulation time step is set to 1 ps. (a) Switching probability of "from $m_z = -1$ to $m_z = +1$" while applying the current along the y-axis; (b) Non-switching probability ($m_z$ remains at $-1$) while applying the current along the x-axis.

It is also worth noting that a large damping constant benefits the robustness of the switching. Figure 3(a) shows the switching probability as a function of the current density at various damping constants. The robustness of the magnetization switching is improved by increasing damping constant. This improvement is attributed to the fact that the higher damping constant can accelerate the energy dissipation and drives the magnetization towards the energy minimum ($+z$ or $-z$) more quickly. Furthermore, enhancing the damping constant only causes negligible increase in the critical switching current density (see Figure 3(b)), as the SOT mainly overcomes the anisotropy field rather than the damping torque. Therefore increasing damping constant is an efficient solution to optimize the proposed switching mechanism. Fortunately, in practical experiments a large damping constant can be naturally obtained in the HM/FM heterostructure due to the spin pumping effect [40-42].

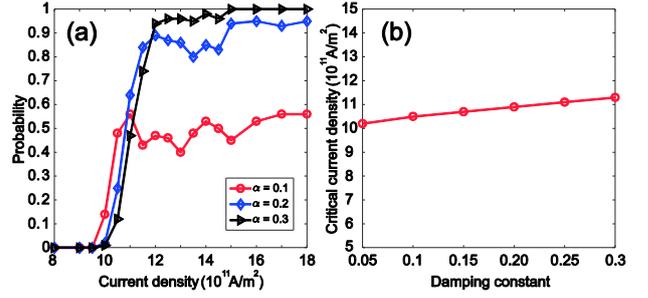

**Figure 3.** (a) Switching probability as a function of current density at various damping constants. The pulse duration is set to 0.5 ns. (b) Dependence of the critical current density on the damping constant.

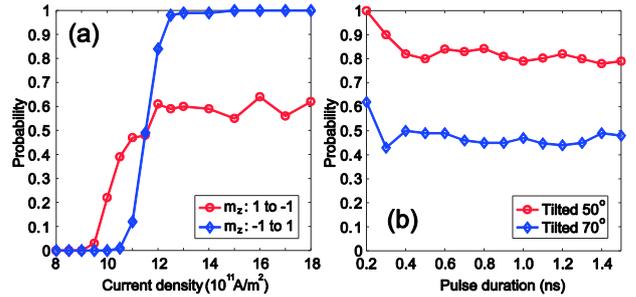

**Figure 4.** (a) Switching probability as a function of current density while $\phi = 60°$. The pulse duration is set to 0.5 ns. (b) Switching probability as a function of pulse duration at various tilted angles. The current density is set to $15 \times 10^{11}$ A/m$^2$.

Next we consider the magnetization dynamics while $\phi \neq 45°$. In this situation, theoretically, the SOT mirror symmetry between the cases $\{+m_0, I_x\}$ and $\{-m_0, I_y\}$ is broken. Thus the switching processes for $+\mathbf{z} \rightarrow -\mathbf{z}$ and $-\mathbf{z} \rightarrow +\mathbf{z}$ are not exactly symmetric. The results of $\phi = 60°$ shown in Figure 4(a) verify the above inference. As can be seen, the $-\mathbf{z} \rightarrow +\mathbf{z}$ switching is more robust than the $+\mathbf{z} \rightarrow -\mathbf{z}$ case. In other words, the magnetization switching shows a preference for $+\mathbf{z}$ direction. Interestingly, as shown in Figure 4(b), the $+\mathbf{z} \rightarrow -\mathbf{z}$ switching probability can be stabilized at a specific value which is insensitive to the pulse duration due to the SOT saturation effect. Furthermore, the stable switching probability can be tuned by changing the tilted angle $\phi$. These properties will benefit the design of true random number generators (TRNGs) with a specified output probability [43, 44].

Figure 4 shows that the $+\mathbf{z} \rightarrow -\mathbf{z}$ switching probability is lower than 1 while $\phi \neq 45°$, which is difficult to be used in the binary memory. Nevertheless, this problem could be solved by increasing the damping constant, based on the theories presented in Figure 3. For instance, while $\phi = 50°$ and the damping constant is increased to 0.3, both $+\mathbf{z} \rightarrow -\mathbf{z}$ and $-\mathbf{z} \rightarrow +\mathbf{z}$ switching probabilities are up to 1. Thus robust bipolar switching can still be achieved with the proposed SOT mechanism, even if $\phi$ is deviated from $45°$.

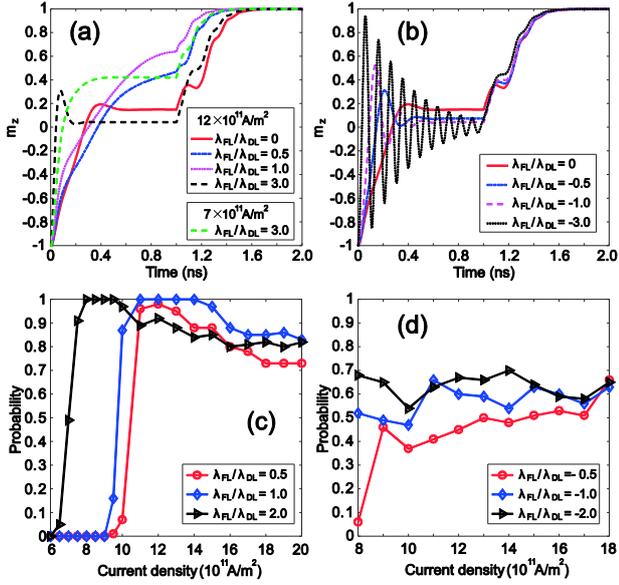

**Figure 5.** (a)-(b) Time evolution of $m_z$ at various values of current densities and $\lambda_{FL}/\lambda_{DL}$. (c)-(d) Switching probability as a function of current density at various values of $\lambda_{FL}/\lambda_{DL}$. In (c)-(d), the pulse duration is set to 1 ns, and the damping constant is set to 0.1.

Finally we study the influence of field-like torque on the magnetization switching. Recent works have demonstrated that a wide range of the $\lambda_{FL}/\lambda_{DL}$ can be obtained [45-47]. Overall, the field-like SOT plays the same role as a transverse magnetic field along $\sigma$. Therefore the magnetic energy landscape can be modified by adjusting the strength of field-like SOT. Assume that the damping-like SOT is absent, the field-like SOT will induce a precessional motion of $m$ and stabilize $m$ at an equilibrium position determined by $H_{eff}$ and $\tau_{FL-SOT}$. Once a positive damping-like SOT is considered ($\lambda_{FL}/\lambda_{DL} > 0$), the damping process of $m$ will be promoted so that the magnetization can be stabilized without the precession (see Figure 5(a)). The equilibrium position can be very close to $\pm z$ by properly tuning the $\lambda_{FL}/\lambda_{DL}$ and current density. Therefore the robustness of the magnetization switching is expected to be improved, which has been validated by simulation results of Figure 5(c). Here we set the pulse duration to 1 ns to ensure that the SOT effect saturates in all cases. As can be seen, even if the damping constant is not strong enough ($\alpha = 0.1$ in Figure 5(c)), the switching probability can be up to 1 in an appropriate range of current density by introducing a non-zero field-like torque. In contrast, in the absence of field-like torque, the robust switching cannot be achieved with a low damping constant (see Figure 3(a)). Note that the switching probability starts to decrease while the current density becomes much larger, because in this case stronger transverse SOT effective field drives the magnetization closer to in-plane direction, which is an unstable position for the switching of perpendicular magnetization.

For larger $\lambda_{FL}/\lambda_{DL}$ (e.g. = 3), strong transverse SOT effective field could lead to a little precession of magnetization, as shown in the black dashed curve of Figure 5(a). This precession can be eliminated by decreasing the current density, as shown in the green dashed curve of Figure 5(a). In other words, the switching current density can be decreased by enhancing the field-like torque, in agreement with results of Figure 5(c). Consider a 80nm-long × 40 nm-width × 3 nm-thick heavy-metal tungsten, the resistivity is around 200 μΩ·cm, the switching current density for 1 ns pulse can be decreased to $6 \times 10^{11}$ A/m$^2$ in the case of $\lambda_{FL}/\lambda_{DL} = 3$. Then the write energy can be as low as 6.9 fJ. Furthermore, this energy promises to be decreased with the scaling of the device.

However, for a negative field-like torque ($\lambda_{FL}/\lambda_{DL} < 0$), the magnetization precession is induced even if $\lambda_{FL}/\lambda_{DL}$ is very small. The reason is that the damping-like SOT resists the damping torque (see Figure 5(b)). We also confirm that decreasing the current density cannot eliminate this magnetization precession (not shown in Figure 5(b) for clarity). Therefore, the robustness of the magnetization switching in the case of negative filed-like torque is degraded, as shown in Figure 5(d).

## 4. Conclusion

We have proposed a novel scheme, based on which the perpendicular magnetization can be deterministically switched by the current-induced SOT in the absence of magnetic field. The switched final magnetization state is dependent on the paths rather than the directions of the applied current, which is completely different from the existing schemes of magnetization switching. Therefore our scheme provides a new freedom for developing novel spintronics memories or logic devices. The switching probability was investigated by considering the thermal fluctuation. It was demonstrated that the robust switching can be achieved at a wide operation window of the applied current pulse. The precise control of pulse amplitude or pulse duration is not required. Furthermore, we found that the performance of the magnetization switching can be improved by increasing the damping constant or adjusting the field-like torque strength. These solutions are practically feasible in the SOT devices with HM/FM bilayer structure.

### Acknowledgements

This work was partly supported by the National Natural Science Foundation of China under Grants 61704005, 61627813, 61571023, the International Collaboration Project B16001, and the National Key Technology Program of China 2017ZX01032101. The authors acknowledge the financial support from the VR innovation platform from Qingdao

**References**

[1] Ikeda S, *et al*. 2010 *Nature Mater*. **9** 721.
[2] Tryputen L, *et al*. 2016 *Nanotechnology* **27** 185302.
[3] Wang M, *et al*. 2018 *Nature Comm*. **9** 671.
[4] Slonczewski J C 1996 *J. Magn. Magn. Mater*. **159** L1.
[5] Berger L 1996 *Phys. Rev. B* **54** 9353.
[6] Miron I M, *et al*. 2011 *Nature* **476** 189.
[7] Liu L, Pai C F, Li Y, Tseng H W, Ralph D C, and Buhrman R A 2012 *Science* **336** 555.
[8] Cubukcu M, *et al*. 2014 *Appl. Phys. Lett*. **104** 042406.
[9] Yu Z, *et al*. 2018 *Nanotechnology*, **29** 175404.
[10] Fukami S, *et al*. 2016 *Nat. Nanotech*. **11** 7.
[11] Lee S W and Lee K J. 2016 *Proc. IEEE* **104** 1831.
[12] Wang Z, *et al*. in *Proceedings of 2018 IEEE International Symposium on Circuits and Systems*, Florence, Italy, 27-30 May, 2018.
[13] Fukami S, Zhang C, Gupta S D, Kurenkov A, and Ohno H. *Nat. Mater*. 2016 **15** 535.
[14] Oh Y W, *et al*. *Nat. Nanotech*. 2016 **11** 878.
[15] Fan Y, *et al*. 2014 *Nat. Mater*. **13** 699.
[16] Mahendra DC, *et al*. 2018 *Nat Mater*. **17** 800.
[17] Nguyen Huynh Duy Khang, *et al*. 2018 *Nat. Mater*. **17** 808.
[18] Yu G, *et al*. 2014 *Nat. Nanotech*. **9** 548.
[19] You L, *et al*. 2015 *PNAS*. **112** 10310.
[20] Lau Y C, *et al*. 2016 *Nat. Nanotech*. **11** 758.
[21] van den Brink A, *et al*. 2016 *Nat. Comm*. **7** 10854.
[22] Wang Z, Zhao W, Deng E, Klein J O, Chappert C. 2015 *J. Phys. D: Appl. Phys*. **48** 065001
[23] van den Brink A, *et al*. 2014 *Appl. Phys. Lett*. **104** 012403.
[24] Cai K, *et al*. 2017 *Nat. Mater*. **16** 712.
[25] Deng J, *et al*. 2018 *Appl. Phys. Lett*. **112** 252405.
[26] Wang X, *et al*. 2018 *Adv. Mater*. **30** 1801318.
[27] Lee J M, *et al*. 2018 *Nano Lett*. **18** 4669.
[28] Legrand W, Ramaswamy R, Mishra R, and Yang H 2015 *Phys. Rev. Appl*. **3** 064012.
[29] Wang M, *et al*. 2018 *Nature Electronics*, **1** 582.
[30] N. Sato, et al. 2018 *Nature Electronics*, **1** 508.
[31] Kazemi M, Rowlands G E, Shi S, Buhrman R A, and Friedman E G 2016 *IEEE Trans. Electron Devices*. **63** 4499.
[32] Luo Z Y, Tsou Y J, Dong Y C, Lu C, and Liu C W. in *Proceedings of 2018 Non-Volatile Memory Technology Symposium (NVMTS)*, Sendai, Japan, 22-24, October, 2018.
[33] Safeer C K, *et al*. 2016 *Nature Nanotechnology* **11** 143.
[34] Wang Z, *et al*. 2018 *IEEE Electron Device Lett*. **39** 343.
[35] Choday S H, *et al*. 2014 *IEEE Electron Device Lett*. **35** 1100.
[36] Garello K, *et al*. 2014 *Appl. Phys. Lett*. **105** L1.
[37] Zhang C, *et al*. 2015 *Appl. Phys. Lett*. **107** 189.
[38] Mikuszeit N, *et al*. 2015 *Phys. Rev. B* **92** 144424.
[39] Brown W F Jr 1963 *Phys. Rev*. **130** 1677.
[40] Couet S, *et al*. 2017 *Appl. Phys. Lett*. **111** 152406.
[41] Lee D J, *et al*. 2018 *Phys. Rev. Appl*. **10** 024029.
[42] Berger A J, *et al*. 2018 *Phys. Rev. B* **98** 024402.
[43] Zhang T, *et al*. 2017 *Nanotechnology*, **28** 455202.
[44] Kim Y, *et al*. 2015 *IEEE Magn. Lett*. **6** 3001004.
[45] Kim J, *et al*. 2013 *Nature Mater*. **12** 240.
[46] Akyol M, *et al*. 2015 *Appl. Phys. Lett*. **106** 032406.
[47] Ou Y, Pai C F, Shi S, Ralph D C, and Buhrman R A 2016 *Phys. Rev. B* **94** 140414.